\documentclass[showpacs,preprintnumbers,amsmath,amssymb,preprint]{revtex4}
\usepackage[final]{graphicx}
\usepackage[english]{babel}
\begin{document}
\title{Jet quenching in high energy nuclear collisions and quark-gluon plasma.}
\author{Yu.A. Tarasov}
\email{tarasov@dni.polyn.kiae.su} \affiliation{
 Russian Research
Center ''Kurchatov Institute'', 123182, Moscow,  Russia }
\date{\today}

\begin{abstract}
We investigate the energy loss of quark and gluon jets in
quark-gluon plasma produced in the central Au+Au collisions at
RHIC energy. We use the physical characteristics of initial and
mixed phases found in the effective quasiparticle model
\protect\cite{5}. We also take into account the possibility of hot
glue production at the first stage.  We calculate the suppression
of $\pi^0$ spectrum, which is caused by the energy loss of gluon
and quark jets. We compare this suppression with data reported by
PHENIX. We find that the suppression is described correctly  by
the quasiparticle model with decrease of thermal gluon mass and
effective coupling  in the region of phase transition (at $T \to
T_c$ from above). Thus it is possible to investigate the structure
of phase transition with the help of hard processes. We also show,
that the energy loss at SPS energy is too small to be observable.
This is caused, in fact, by too low initial temperature of the
plasma phase at this energy.
\end{abstract}

\pacs{12.38.Mh, 24.85.+p, 25.75.-q}

\maketitle

\section{Introduction    \label{sec1}}

Energy loss of high energy quark and gluon jets in relativistic
A+A collisions leads to jet quenching and thus probes the
quark-gluon plasma~\cite{1}. Recently the considerable suppression
of hadron spectra at moderately high $p_{\perp} <6$ GeV was found
in the central Au + Au collisions, especially for $\pi^0$
production~\cite{2}. This suppression can be explained by the
energy loss of the moderate $p_{\perp}\le 10$ GeV gluons jets.
This is one of the major new results at RHIC energy, which has not
been observed before at lower SPS energy. The energy loss was
investigated in the work~\cite{3} in various orders in opacity
$L/\lambda_g$(where $L$ is nuclear radius and $\lambda_g$ is the
gluons mean free path). It was shown that these series is strongly
dominated by the first term. It was also shown in Ref.\ \cite{4}
that finite kinematic boundaries decrease the energy loss at
intermediate jet energies as compared to the asymptotic limit.
Recently, in paper \cite{5}, the physical characteristics of
initial plasma phase and also of mixed phase at SPS and RHIC
energies (i.e. the values $T_0$, $\tau_0$, $\tau_c$) were
investigated  on the basis of the quasiparticle model and
isentropic expansion. In the present work we use these initial
conditions for investigation of the energy loss of high energy
gluon and quark jets at RHIC energy. In these studies we take into
account also the possibility of two stages of equilibrium, i.e.
production of the hot glue at first stage~\cite{6}. The
calculation of energy loss is important for investigation of the
phase transition structure of quark-gluon plasma into hadrons. In
the Refs.\ \cite{7,8} the phenomenological parametrization of
coupling constant $G(T)$ was used in accordance with new lattice
data. When approaching to phase transition point from above, the
decrease of thermodynamic values is caused, according to $SU(3)$
gauge theory, by increase of the thermal gluon mass and also of
the coupling constant $G(T)$. In these conditions the
applicability of perturbation theory is questionable. In
phenomenological model of confinement~\cite{9} the decrease of
thermal gluon mass $m_{g}(T)$ at $T\to T_c$ from above is assumed,
which is connected with decrease of effective coupling strength
$G(T)$. It can be understood, as the more and more gluons  become
confined and form heavy glueballs with decrease of T, and the
effective glueball exchange interaction between gluons are
reduced. However, in this case the entropy density $s(T)$ will
exceed the lattice entropy because of domination of light masses
near $T_c$. This difference may be accounted for in quasiparticle
model by modification of the number of effective degrees of
freedom in thermodynamic functions:
\begin{equation}
\label{eq.1} g_g \to C(T)g_g.
\end{equation}
The explicit value $C(T)$ may be estimated as ratio of the lattice
entropy and the quasiparticle entropy density $s_{g}(T)$ with decrease of the
mass $m_{g}(T)$. At $T\gg T_c$ we have $C(T)\simeq 1$, and near
$T\sim T_c$ we have $C(T)< 1$. The value $C(T)$ for gluons has the form
\cite{9}:
\begin{equation}
\label{eq.2} C(T,T_c) = C_{0}(1+\delta_c
-\frac{T_c}{T})^{\beta_c}\,,
\end{equation}
where $C_0 \simeq$ 1.25, $\delta_{c} \simeq$ 0.0026, $\beta_{c}
\simeq$ 0.31.  The thermal mass $m_{g}(T)$ can be parameterized
well by formula:
\begin{equation}
\label{eq.3} m_{g}(T) = (\frac{N_c}{6})^{1/2} G_{0}T(1+\delta
-\frac{T_c}{T})^{\beta},
\end{equation}
where $\beta \simeq$ 0.1, $\delta \simeq$ 10$^{-6}$. It possible
to vary some parameters, for example $\beta \simeq 0.05, \delta
\simeq$ 10$^{-7}$ are reasonable~\cite{9}. The form (\ref{eq.3})is
analogous to behaviour of Debye mass $m_D$, extracted from lattice
data~\cite{8,9}. The value $G_0$ can be determined by asymptotic
value of the thermal mass, chosen from coincidence of lattice and
perturbative masses at $T\simeq 3T_c$ \cite{7,8}. We had found for
pure gluons: $G_0 \simeq$ 1.9 \cite{5}.

The relations (\ref{eq.2}), (\ref{eq.3}) with above-mentioned
parameters give good description of $SU(3)$ lattice data for
entropy $s$, energy density $\epsilon$, and pressure $p$ for $T$
close to $T_c$. It is possible to extend the effective
quasiparticle model to a system with dynamic quarks, with
analogous effective coupling $G(T)$, and the function $C(T,T_c)$
with some variation of parameters~\cite{9}. One can show that the
same constant $G_{0}\simeq$ 1.9 describes the lattice data for 2
and 2+1 flavors (for example, the lattice pressure~\cite{5,8}.

It is interesting to note, that the massive constituent quarks ($m_q$ and
$m_s$) appears also due to the decrease of number of degrees of freedom in the
presence of octet of pseudogoldstone states. This is a consequence of
conservation of the entropy and of the  number of net nucleons. It can also be
demonstrated that hadrons (in the hadronic part of the mixed phase) appears
with the same effective number of degrees of freedom \cite{5}.

It can be shown, that such picture of phase transition (with
decrease of effective coupling strength $G(T)$ near $T \simeq T_c$)
gives reasonable quantitative description of the jet quenching and of
suppression of the hadronic spectra in the central Au +Au collisions at
RHIC energy. However, the use of the ``running'' coupling $\alpha_{s}(T)$
in the perturbative decomposition of the thermodynamic values in plasma
leads to too large energy loss of jets, which disagrees with
experimental data for suppression of the hadrons with large $p_{\perp}$ at RHIC
energy.

In the Ref.\ \cite{5} we do not find a noticeable difference for
meson and baryon spectra in the ordinary perturbative theory with
the running coupling $\alpha_{s}(T)$ and also in model with
phenomenlogical parametrization of coupling $G(T)$ \cite{7,8} in
comparison with the effective quasiparticle model. However, the
ordinary perturbative model disagrees with $SU(3)$ lattice data in
the region of phase transition. Thus the spectra of particles
apparently weakly depend on the character of phase transition.
Therefore the investigation of jet quenching represent significant
interest.

In Sec.\ \ref{sec2} we calculate the energy loss of the high energy
gluon and quark jets in quark-gluon plasma at RHIC energy. In these
calculations we use the initial conditions in plasma at SPS and
RHIC energies, which were found in the quasiparticle model~\cite{5}. We also
take into account here the production of hot glue at the first
stage. We show that the energy loss at SPS energy is too small. We
show also, that in the perturbative theory the energy loss of gluon
jets is too large.

In Sec.~\ref{sec3} we calculate the suppression of $\pi^0$ at
moderate $p_{\perp}$ ( $3\le p_{\perp}\le6$) GeV in the central
Au+Au collisions. We take into account the jet quenching and the
parton shadowing factor in nucleus. We show that suppression of
the $\pi^0$ spectra at $P_{\perp}^{\pi^0}\ge 3$ GeV/c does not
contradict the experimental data at RHIC energy.

\section{Energy loss of high energy jets in quark-gluon plasma   \label{sec2}}

In the effective quasiparticle model~\cite{5} we have investigated the initial
condition and the evolution of the plasma stage, where there is equilibrium for
both quarks and gluons. We have found for RHIC case the values $T_0 \simeq
216.3-219.6$ MeV, $\tau_{0}\simeq 2.22-2.18$ fm and $\tau_{c}\simeq 6.34-6.52$
fm. We have also found corresponding values for SPS energy $T_0 \simeq 175$
MeV, $\tau_{0}\simeq 3.28$ fm, $\tau_{c}\simeq 4.1$ fm, so, here we have the
short plasma stage. The production of a more hot glue plasma at the first stage
is caused by the relatively large $gg$ cross section in comparison with the
$qg$ and $qq$ cross section of Ref.\ \cite{6}. In the lowest order, the matrix
elements $M^2$ in formula:
\begin{equation}
\label{eq.4}
\frac{d\sigma}{dt} = \frac{\pi\alpha_s^2}{s^2} M^2
\end{equation}
at large angles are related thus as \cite{10}:
\begin{equation}
\label{eq.5} M^2_{gg\to gg}/M^2_{qg\to qg}/M^2_{qq\to qq} = 30.4/
5.4/2.2
\end{equation}
i.e.\ $gg$ scattering is most important. The small angle scattering leads to
divergent cross-sections, which however are finite in $QGP$ due to the finite
``Debye mass'' $t_{\min} = g^{2}T^2$. The cross section at large angles is
defined by formula:
\begin{equation}
\label{eq.6} \sigma_{gg}^{large-angle}\simeq \frac{30.4 \pi
\alpha_s^2}{2\bar s}\simeq \frac{95.46 \alpha_s^2}{36 T^2}\simeq
2.65\frac{\alpha_s^2}{T^2}.
\end{equation}
The cross section at small angles has the form:
\begin{equation}
\label{eq.7} \sigma_{gg}^{small-angle}\simeq \frac{9}{2}
\frac{\pi\alpha_s^2}{4\pi T^{2}\alpha_s} =\frac{9}{8}
\frac{\alpha_s}{T^2}.
\end{equation}
The effective scattering rate $1/\tau_g$ is determined by a sum of
large and small angle values:
\begin{equation}
\label{eq.8} \frac{1}{\tau_g} = n_{g}(2.65\frac{\alpha_s^2}{T^2} +
\frac{9}{8} \frac{\alpha_s}{T^2}).
\end{equation}
Here the value $\alpha_s$ is determined by the effective coupling
constant $\alpha_s = \frac{G^2}{4\pi}$, where:
\begin{equation}
\label{eq.9} G(T,T_c) = G_{0}(1+\delta - \frac{T_c}{T})^{\beta}
\end{equation}
and $G_{0}\simeq 1.9$ (see Eq.\ (\ref {eq.3}))

The net gluon density in effective quasiparticle model is:
\begin{equation}
\label{eq.10}
 n_{g}(T) = \frac{16 T^3}{2\pi^2}
\int\limits_0^{\infty}dx\, \frac{x^{2} C(T,T_c)}{e^{\sqrt{x^2 +
\frac{m_g^2}{T^2}} - 1}}\,,
\end{equation}
where
\begin{equation}
\label{eq.11} \frac{m_g^2}{T^2} = \frac{1}{2} \left[G_{0}\left(1+
\delta - \frac{T_c}{T}\right)^{\beta}\right]^2\,.
\end{equation}

To estimate the initial temperature of the hot glue we use Bjorken model
\cite{11}. The value $dN/dy$ (number of charged and neutral particles per unit
central rapidity) for $Au + Au$ collisions can be described by formula
\begin{equation}
\label{eq.12} \left(\frac{dN}{dy}\right)_{y=0} \simeq 0.8 \ln\sqrt{s}
A_{Au}^{1.11} \,.
\end{equation}
This value agrees with experimental value 1374 at $\sqrt{s}=130$ GeV.

In this paper we make the following assumptions: at the first stage we have the
hot glue in equilibrium and nonequilibrium quarks. The total entropy of gluons
and such quarks (at temperature $T_g$) is less than the entropy of gluons and
\textit{all} quarks at more a later stage (at above mentioned temperature
$T_0$). With expansion of the system the temperature  decreases and the entropy
of quarks and gluons decreases (if to assume that all the quarks are
nonequilibrium ones). However, in reality, some part of quarks achieves
equilibrium and entropy increases. With subsequent decrease of T the number of
equilibrium quarks increases, and near $T_0$ we have the equilibrium of both
gluons and quarks. After that the cooling becomes isentropic.

One can estimate the fraction of gluons and sea $u,d,s$ quarks
(and antiquarks) from the structure function \cite{12}, if to
imagine the system moving backward in time from many secondaries
into hot partonic plasma. This approach has to normalize parton
multiplicity to the total entropy, later observed as multiplicity
of secondaries. We assume its conservation at later stage (for $T
\le T_0$) and its additional production at previous
stage~\cite{6}. For average number of partons (for example, gluons
in nucleon) we use formula~\cite{13}:
\begin{equation}
\label{eq.13} N_g=\int\limits_{m_{char}/|p|}^{1} dx \, g(x,Q),
\end{equation}
where $g(x,Q)$ is structure function, $m_{char}\sim m_{\rho}$ is characteristic
hadrons mass, and $|p|$ is nucleon momentum. For RHIC energies we have
$m_{char}/|p| = 2m_{char}/\sqrt{s}\approx 0.012$.

Similarly, the number of the sea quarks is
\begin{equation}
\label{eq.14} N_{sea} = 2 \int \limits_{m_{char}/|p|}^{1}dx
\,(u_{sea}+d_{sea}+s_{sea}) \,.
\end{equation}
This gives estimate $N_{sea}/N_g \approx 0.44$. This correspond to
known fact,that gluons dominate in structure function at small x:
 the number of gluons is more than twice larger than the number
of sea quarks and antiquarks. The characteristic average value
$x_{\min}\simeq 0.012$ is close to value $x_{\min}\simeq 0.01$ in
the works of other authors (for example E.V.Shuryak). These
meanings are reasonable for estimation of the hot glue temperature
$T_g$, since partons with $x < x_{\min}$ carry only a small part
of the total momentum. For estimation of the number of gluons  in
central region of rapidity we use the relation (\ref{eq.12}):
\begin{eqnarray}
\label{eq.15} && \left(\frac{dN_g}{dy} + \frac{dN_{sea}}{dy} +
\frac{dN_{val}}{dy}\right)_{y=0} \simeq  \nonumber \\ &\simeq& \frac{dN_g}{dy}
+ 0.44 \frac{dN_g}{dy} +\frac{dN_{val}}{dy} \simeq 1374\,,
\end{eqnarray}
where $N_{val} = 3(N - \bar N) \simeq 49$ (where $N - \bar N \simeq 16.3$
\cite{5} is the number of net nucleons for RHIC in central region of rapidity).

Hence, the estimate for gluons is:
\begin{equation}
\label{eq.16} \left(\frac{dN_g}{dy}\right)_{y=0} \simeq 920.
\end{equation}
Taking into account Eq.\ (\ref{eq.8}), we have the relation for determination
of the hot glue initial temperature $T_g$:
\begin{equation}
\label{eq.17}
n_{g}\tau_{g} = \frac{dN_g}{dy \pi R_{Au}^2} \simeq
12.5 (m_\pi)^2
 = \frac{T^2}{2.65 \alpha_{s}^{2}(T) + 1.125 \alpha_{s}(T)} \,.
\end{equation}
Here the physical values are in the units of $m_{\pi}=139$ MeV.
The function $\alpha_{s}(T)$ is determined by Eq.\ (\ref{eq.9}).
We find from Eq.\ (\ref{eq.17}):
\begin{equation}
\label{eq.18} T_{g}\simeq 345\, \text{MeV}
\end{equation}
 From Eq.\ (\ref{eq.10}) the initial gluon density is:
\begin{equation}
\label{eq.19}
 n_{g}(T_g) \simeq 19.42 \; m_{\pi}^3.
\end{equation}
Thus the time required to achieve the equilibrium for gluons is
\begin{equation}
\label{eq.20}
 \tau_{g}\simeq \frac{12.5}{n_{g}(T_g)}\simeq
\frac{0.64}{m_{\pi}} \simeq 0.91 \; \text{fm/c}\,.
\end{equation}
The same time for quarks and gluons found in Ref.\ \cite{5} is $\tau_{0} \simeq
1.54/m_{\pi} \simeq 2.18$ fm/c. This agrees with estimate in Ref.\ \cite{6}.

In the region $T_0 <T\le T_g$ the value $\tau_g$ we estimate also by Eq.\
(\ref{eq.8}). Such estimates are reasonable, as in the effective quasiparticle
model the coupling constant $\alpha_{s}(T)$ decreases with decrease of
temperature, and Eqs.\ (\ref{eq.6})--(\ref{eq.8}) in the lowest order are a
good approximation. In fact, in the effective quasiparticle isentropic model we
have the initial conditions in plasma: $T_0 \simeq 219.6$ MeV and $\tau_0
\simeq 1.54 /m_{\pi}$~\cite{5}. However, from Eq.\ (\ref{eq.8}) at $T$ = 219.6
MeV and $n_g = 3.7 m_{\pi}^3$ (for the case of equilibrium quarks \cite{5}, see
Table \ref{Tab.1}) we have $n_{g}\tau_{g}= 5.67 m_{\pi}^2$, which gives the
same value $\tau_{g} = 1.53/m_{\pi}$. Thus Eq.\ (\ref{eq.8}) provides matching
of values $\tau$ for the two mentioned regions.

The values $\tau$ for different values of $T$ are shown in Table~\ref{Tab.1}.
In this table we also give the estimates of equilibrium light quark density in
the region $T_0 < T < T_g$, where the total equilibrium of gluons and quarks is
not achieved (we assume that in the initial state of the hot glue at $T_g =
345$ MeV all the quarks are nonequilibrium ones)

These estimates can be done in the following way:

1. We calculate the number of gluons $N_g = n_{g}\tau_{g}\pi R_{Au}^2$ in the
region of the hot glue $T_0 < T < T_g$. The number of the nonequilibrium quarks
$N_{q+s}$ we estimate using Eq.\ (\ref{eq.15}) and subtracting the value $N_g$.

2. For these quarks we calculate the nonequilibrium entropy
$S_{q+s}$ in the Boltzmann approach (the expected difference for
calculations with fermions is about $\sim 5\%$). We calculate also
the entropy $S_g$ for equilibrium gluons. We assume, that the
total entropy $S_g + S_{q+s}$ increase smoothly (almost linearly
like $N_{q+s}(T)$) to the equilibrium entropy $S_{0}(T_0)\simeq$
6300 with decrease of $T \to T_0$~\cite{5}.Using this assumption,
we estimate the equilibrium addition from quarks to the total
entropy density , and equilibrium addition $n_{q}(T)$ to quark
density in the region $T_0 < T < T_g$.  The equilibrium densities
$n_{g}(T)$ and $n_{q}(T)$ in Table~\ref{Tab.1} (at $T_c\le T \le
T_0$) we find, using the formulas of effective quasiparticle
model~\cite{5}.

The dominating first order radiation intensity distribution for expanding
plasma is given in Ref.\ \cite{4}:
\begin{eqnarray}
\label{eq.21} &&\frac{dI}{dx} = \frac{9 C_{R} E}{\pi^2}
\int\limits_{z_0}^{\infty}dz\,\rho (z)
\int\limits_{|k|_{\min}}^{|k|_{\max}} d^{2} {\textbf k}\ ,
\alpha_{s} \nonumber \\ & &  \int\limits_0^{q_{max}} d^{2}
{\textbf q} \frac{\alpha_{s}^2}{[{\textbf q}^{2}+{\mu}^{2}(z)]^2}
\frac{{\textbf k}
{\textbf q}}{{\textbf k}^{2}( {\textbf k} - {\textbf q})^2}\times \nonumber \\
&\times&\left[1 - \cos\frac{( {\textbf k} - {\textbf
q})^{2}(z-z_0)}{2x(1-x)E}\right] \,,
\end{eqnarray}
where $E$ is jet energy, and $C_{R}$ is color factor of jet ($C_{R} = N_c$ for
gluons). Here we take into account substitution $x \to x(1-x)$ for $x \to 1$
\cite{14}.   It is assumed that quark-gluon plasma can be modeled by well
separated color-screened Yukawa potentials. The upper kinematic bound of medium
induced momentum transfer is $|q|_{\max}\approx \sqrt{3\mu(\tau)E}$. The
transverse momentum ${\textbf k}_{\perp}^2$ is connected with gluon emission
distribution from parton jet in the absence of a medium. The kinematic bound on
the transverse momentum is ${\textbf k}_{\max}^2= \min[4E^{2}x^2,4E^{2}x(1-x)]$
and ${\textbf k}_{\min}^2 = \mu^{2}(\tau)$  for  gluons with light cone
momentum fraction $x$ (where $\mu^{2}(\tau)=4\pi\alpha_{s}T^{2}(\tau)$).

The value $z=\tau$ is limited by thickness L of target, but as a matter of
fact, by end of plasma phase and by beginning of mixed phase $\tau_c$, i.e. by
duration of plasma phase $\tau_0\le\tau\le\tau_c$. The value $\tau_c$ does not
exceed the radius  $R_{Au}$ (or it is close  to it). The value $\rho(\tau)$ is
the gluon or quark density at time $\tau$ along the jet path (i.e. $\rho = n_g$
or $n_q$). The energy loss $\Delta E$ is defined by integration of $dI/dx$ in
Eq.\ (\ref{eq.21}) over $x$. The value
\begin{eqnarray}
 \label{eq.22}
 &&I_{1}(E,\tau(T)) = \int\limits^{1}_{0} dx
 \int\limits_{|{\textbf k}|_{\min}}^{|{\textbf k}|_{\max}}d^{2}{\textbf
k}\,\alpha_{s}(\tau)
 \nonumber \\ & &
 \int\limits_{0}^{|{\textbf q}|_{\max}}\frac{d^2 {\textbf q}\,
 \alpha_{s}^{2}(\tau)}{[{\textbf q}^2+\mu^2(\tau)]^2}
 \frac{{\textbf  k}{\textbf q}}{{\textbf k}^2({\textbf k}- {\textbf q})^2}
\times
 \nonumber \\ &\times&
 \left[1-\cos{\frac{({\textbf k}- {\textbf q})^2\tau}{2x(1-x)E}} \right]
\end{eqnarray}
in the finite kinematic bounds is calculated by Monte Carlo method for values
of $\tau(T)$ in the region $\tau_{g}\le\tau\le\tau_c$ (for RHIC energy) and for
several values of $E$. In the calculations of complete energy loss $\Delta E$
the scattering of gluons (and quarks) by gluon and quark potentials in plasma
medium  should be taken into account for every value of $E$. For example, the
fraction of gluon scattering on gluon potential in medium is defined by
formula:
\begin{equation}
\label{eq.23} \frac{n_{g}\sigma_{gg}}{n_{g}\sigma_{gg}+n_{q}\sigma_{gq}} =
\frac{1}{1 + \frac{n_{q}\sigma_{gq}}{n_{g}\sigma_{gg}}}\,,
\end{equation}
where $\sigma_{gg}$ is the cross section for gluon scattered on gluon
potential, and $\sigma_{gq}$ is correspondingly cross section for gluon
scattering on quark potential. Also the fraction of gluon scattering on quark
potential is $(1+\frac{n_{g}\sigma_{gg}}{n_{q}\sigma_{qg}})^{-1}$. Taking into
account the decrease of the color factor we have $\sigma_{qg}/\sigma_{gg}=
4/9$~\cite{3}.

The complete energy loss $\Delta E$ of gluon jet is determined by
formula:
\begin{eqnarray}
\label{eq.24} \Delta E_{g}(E) &=&
\frac{9C_{R}E}{\pi^{2}\mu^2}\int\limits_{\tau_g}^{\tau_c}d\tau\,
I_{0}(E,\tau)\frac{n_g}{1+\frac{4n_q}{9n_g}} + \nonumber \\ &+&
\frac{4C_{R}E}{\pi^{2}\mu^2}\int\limits_{\tau_g}^{\tau_c}d\tau\,I_{0}(E,\tau)
\frac{n_q}{1+\frac{9n_g}{4n_q}}\,,
\end{eqnarray}
where the value $I_{0}(E,\tau)$ corresponds to $I_{1}(E,\tau)$ in variable
$|k|/\mu$ and $|q|/\mu$. Similarly, the energy loss of quark jet is determined
by formula:
\begin{eqnarray}
\label{eq.25}
\Delta E_{q}(E) &=&
\frac{4C_{F}E}{\pi^{2}\mu^2}\int\limits_{\tau _g}^{\tau_c}d\tau \,
I_{0}(E,\tau)\frac{n_g}{1+\frac{4n_q}{9n_g}} + \nonumber \\
&+& \frac{16C_{F}E}{9\pi^{2}\mu^2}\int\limits_{\tau_g}^{\tau_c}d\tau\,
I_{0}(E,\tau)\frac{n_q}{1+\frac{9n_g}{4n_q}}\,,
\end{eqnarray}
where $C_F$ is the color factor of quark: $C_{F} = \frac{N_c^{2}-1}{2N_c}$.

However it should be noted, that for the well-separated color
screened Yukawa potentials the condition $\lambda =
\frac{1}{\sigma \rho} \gg 1/\mu$ must be implemented (here
$\lambda$ is the mean free path of partons, and $\mu$ is the color
screening parton mass). This condition is well realized for parton
scattering on gluon potential, where $\mu_g^2\simeq
4\pi\alpha_{s}T^2$, and we have $\lambda_{g,q}\mu_{g}\gg$ 1. But
this condition is not realized for partons scattering on the quark
Yukawa potential, where $\mu_q^2 = 4\pi\alpha_{s}T^{2}/6$, and we
have here $\lambda_{g,q}\mu_{q}\le$ 1. Therefore the estimation of
partons energy loss $\Delta E$ on gluon Yukawa potential is the
most real in this model. These energy loss $\Delta E$ of gluon jet
we give in Table~\ref{Tab.1} for example at E = 10 GeV. But
calculating also the energy loss on quark potential by formulas
(\ref{eq.24} - \ref{eq.25}), we find the decrease of $\Delta E$ no
more than 10\%.

\begin{figure}[ht]
\centering \mbox{\includegraphics*[scale=0.8]{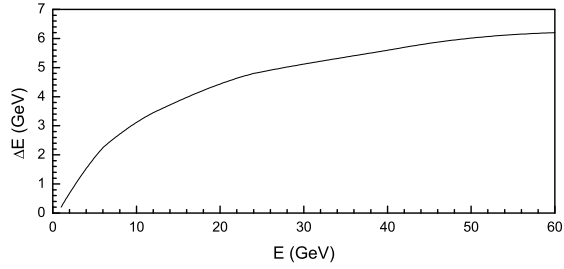}}
 \caption{The energy loss of gluon jet in quark-gluon
plasma at RHIC energy. \label{Fig.1}}
\end{figure}

\begin{figure} [ht]
\centering \mbox{\includegraphics*[scale=0.8]{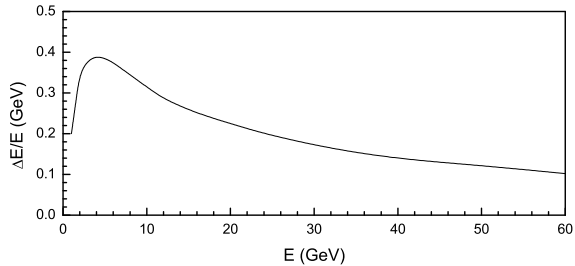}}
 \caption{The relative energy loss of gluon jet in
quark-gluonf plasma. \label{Fig.2}}
\end{figure}

  The complete energy loss $\Delta E$ is calculated by
numeral integration over $\tau$. We have $\Delta E_{g}\simeq 3.15$
GeV at $E = 10$ GeV. The complete energy loss of gluon jets
$\Delta E_{g}(E)$ is shown in Fig.\ \ref{Fig.1} opposite $E$ (Fig.
\ref{Fig.2} shows the relative energy loss $\Delta E(E)/E$).

\begin{table}
\caption{ The values of various physical quantities for expanding plasma at
RHIC energy and energy loss of gluon jet $\Delta E$. \label{Tab.1} }
\begin{tabular}{ccccccc} \hline \hline
$T$ (MeV)& $\tau$ ($m_{\pi}^{-1}$)  & $\alpha_s$ & $I_0$ & $n_g$
($m_{\pi}^{3}$) & $n_q$ ($m_{\pi}^{3}$) & $\Delta E$ (GeV)
\\ \hline
345 & 0.645 & 0.267 & 0.092 & 19.42 &  0 & 2.367
\\
325 & 0.70 & 0.265 & 0.088 & 15.95 &  4.41   & 2.097
\\
300 & 0.79 & 0.262 &  0.085  & 12.22 & 4.71    & 1.855
\\
275 & 0.908 & 0.26 &  0.083  & 9.09  & 4.45     & 1.585
\\
250 & 1.075 & 0.257 &  0.081  & 6.5   &   4.21   & 1.38
\\
225 & 1.34 & 0.252 &   0.079   &4.4   &   3.50    & 1.13
\\
219.6 & 1.54 & 0.247 & 0.078   & 3.7  & 6.1  & 1.02
\\
210  & 1.74  & 0.241  & 0.077   & 3.12  & 5.134 & 0.964
\\
200  & 2.04   & 0.235   & 0.075   & 2.56  & 4.195 & 0.84
\\
190  & 2.44   & 0.229   & 0.073   & 2.045  & 3.32  & 0.778
\\
185   & 2.72   & 0.222  & 0.067   &  1.795 & 2.9   & 0.663
\\
180   & 3.1    & 0.215  & 0.063   &  1.545  & 2.47  & 0.589
\\
175   & 3.7    & 0.201   & 0.055  &   1.286  & 2.03  & 0.48
\\
172   & 4.26   & 0.184   &  0.043  &   1.116  & 1.73  & 0.37
\\
170   &  4.6    & 0.057  &  0.00067 &   1.23  &  1.61  & 0.0213
\\   \hline \hline
\end{tabular}
\end{table}

 Let us consider now the energy loss at SPS energy. For the value
 $dN/dy$ per unit central rapidity (charged + neutral) we have
 ~\cite{5}:
 \begin{equation}
 \label{eq.26}
 \left(\frac{dN}{dy}\right)_{y=0}\simeq 803.
 \end{equation}
 The number of partons from structure function is also obtained
 by Eqs.\ (\ref{eq.13}),(\ref{eq.14}), but now $x_{min}\simeq 0.09-0.1$.
 The fraction of sea quarks is $N_{sea}/N_{g}\simeq 0.58$. Taking into
 account the  number of valent quarks in the central region of rapidity
 $\simeq 57\times 3$ we have now:
 \begin{equation}
 \label{eq.27}
 \left(\frac{dN_g}{dy}\right)_{y=0}\simeq 400.
 \end{equation}
 We use the relation analogous to (\ref{eq.17}) for derivation of the
 hot glue initial temperature $T_g$:
 \begin{equation}
 \label{eq.28}
 \frac{1}{\pi R_{Pb}^2}\left(\frac{dN_g
 }{dy}\right) \simeq 5.1.
 \end{equation}
 It can be shown, however, that there is no solution of equation (\ref{eq.17}) for
 $T_g$ in that case. Apparently, at such low energy the
 hot glue phase is practically indistinguishable from phase of
 total equilibrium(or close to it). In Table \ref{Tab.2} the energy loss $\Delta E(\tau)$ is
 shown at SPS  energy ($E=4$ GeV) for the case of total equilibrium.
 At $E=4$ GeV the total energy loss is $\Delta E \simeq 106$ MeV. Such small energy
 loss is  caused by a short lifetime of the plasma phase at low SPS energy. It should be
 noted, that the energy losses, which are found here, correspond to
 the model of phase transition with decrease of
 effective coupling strength $G(T)$ and mass $m_{g}(T)$ for $T\to T_c$
 from above.

\begin{table}
\caption{The same as Table I but for SPS energy.\label{Tab.2}}
\begin{tabular}{ccccccc} \hline \hline
$T$ (MeV)& $\tau$ ($m_{\pi}^{-1}$)  & $\alpha_s$ & $I_0$ & $n_g$
($m_{\pi}^{3}$) & $n_q$ ($m_{\pi}^{3}$) & $\Delta E$ (GeV)
\\ \hline
175  & 2.4  & 0.201 & 0.061 & 1.28  & 2.2  & 0.212
\\
174  & 2.52 & 0.197 & 0.058  & 1.224 & 2,1  & 0.204
\\
173  & 2.65 & 0.192 & 0.055  & 1.168  & 1.99 & 0.188
\\
172  & 2.8  & 0.184 & 0.049  & 1.11   &  1.88 & 0.168
\\
171  & 2.9 & 0.172 & 0.04   & 1.056  &  1.76 & 0.141
\\
170  & 2.98 & 0.057 & 0.00075 & 1.23  &  1.74 & 0.0091 \\
\hline\hline
\end{tabular}
\end{table}

 The similar problem was considered in the Ref.\ \cite{5} using the
 ordinary perturbation theory for
 powers of running coupling $\alpha_s$ in quark-gluon plasma (up
 to order $O(\alpha_s)$). In this approach the coupling
 $\alpha_s$ increases with $T\to T_c$ above. Although this model
 disagrees with SU(3) lattice data in the region of phase transition
 (i.e. close to $T_c$), we do not find here a noticeable difference
 for spectra of particles in comparison with the effective
 quasiparticle model (i.e.the spectra weakly depend on structure
 of phase transition).

However, let us consider the energy loss at RHIC in this
perturbative model. We use the Eq.\ (\ref{eq.8}) with running
coupling $\alpha_s$~\cite{5} (with $\lambda \simeq 180$ MeV) and
values $n_g$ without quarks and with quarks in the  perturbative
decomposition. We find the initial temperature of the hot glue to
be $T_{g}\simeq 400$ MeV. The initial temperature of the total
equilibrium is $T_0 \simeq 219$ MeV, $\tau_0 \simeq 2.18$ fm and
$\tau_c \simeq 3.73$ fm~\cite{5}. The total energy loss can be
found from corresponding Table, which analogous to Table
\ref{Tab.1}. For example, we find $\Delta E\simeq 13.5$ GeV for
$E=16$ GeV, $\Delta E\simeq 8.5$ GeV for $E=10$ GeV, and $\Delta
E\simeq 3.5$ GeV for E=4 GeV. These energy losses as too high and
disagree with the data for suppression of hadrons with large
$p_{\perp}$ in the central nuclear collisions at RHIC energy. This
disagreement is connected with injustice of perturbation theory in
region of phase transition. In Ref.\ \cite{5} the quasiparticle
model with phenomenological parametrization of coupling constant
$G(T)$ was also considered. In this model $G(T)$ increase for $T
\to T_c$ from above. This model agrees with the new lattice data
and provide also good description of the baryon and meson spectra.
However, it can be shown that in this model we have a paradox
result for energy loss of gluon jet in plasma: the energy loss
exceeds the energy of jet itself (even without taking into account
the effect of the hot glue). This contradiction is caused by too
large value of coupling $G(T)$, especially close to the phase
transition point.

 \section{Suppression of pions with large
 ${p}_{\perp}$ in central Au+Au  collisions \label{sec3}}

The jet quenching reduced the jet energy before fragmentation,
where the jet transverse momentum is shifted by energy loss on to
value $\Delta E(E)$~\cite{15}. To account for this effect we
should replace the vacuum fragmentation function by effective one
$z_{c}^{*}/z_{c} D_{h/c}(z_{c}^{*},\hat Q^2)$, where
\begin{equation}
\label{eq.29} z_{c}^{*} =\frac{z_c}{1-\frac{\Delta E(E)}{E}}\,.
\end{equation}
The invariant cross section of hadron production in central A+A collisions is
given by:
\begin{eqnarray}
\label{eq.30} &&E_{h}\frac{d\sigma_{h}^{AA}}{d^3p} =
\int\limits_0^{b\max}d^2b d^2r \,t_{A}(r) t_{A}(|\textbf  b -
\textbf r|) \nonumber \\ && \sum_{abcd}\int
dx_{a}dx_{b}d^{2}k_{\perp,a}d^{2}k_{\perp,b} \times \nonumber
\\&\times& g_{A}(k_{\perp,a},Q^{2},r) g_{A}(k_{\perp,b},Q^2,|\textbf
b -\textbf  r|) \times \nonumber \\ &\times&
f_{a/A}(x_{a},Q^2,r)f_{b/A}(x_b,Q^2,|\textbf  b - \textbf r|)\times \nonumber
\\&\times&\frac{d\sigma}{d\hat t} \frac{z_{c}^{*}}{z_c}
\frac{D_{h/c}(z_{c}^{*},\hat Q^2)}{\pi z_c}\,.
\end{eqnarray}
Here $t_{A}(r)$ is the nuclear thickness function, $k_{\perp,a}$ and
$k_{\perp,b}$ are the initial transverse momenta of partons, $f_{a/A}$ and
$f_{b/A}$ are parton structure functions. It is usually assumed that
distribution $g_{A}(k_{\perp})$ has a Gaussian form.

It should be noted, that intrinsic $k_{\perp}$ and the transverse momentum
broadening (Cronin effect) are important for final hadron spectra for SPS
energies. However, with the increase of energy the spectra become flatter and
small initial $k_{\perp}$ correspond to small variation of spectra. At RHIC
energy one can neglect the effects of initial $k_{\perp}$ with a good precision
\cite{16}. In this work we do not take into account the intrinsic transverse
momentum at RHIC energy.

For parton distribution of nucleon in nucleus we take into account the parton
shadowing factor $S_{a,b/A}(x,r)$ for spherical nucleus and $S_{a,b/A}(x)$ for
flat disk, for which we take the parametrization used in HIJING
model~\cite{17}. In this model the factor $S_{a/A}(x,r)$ is splitted into two
parts: $S(x,r)=S_{0}(x)- \alpha_{A}(r)S_{1}(x)$, where $\alpha_{A}(r)=
0.1(A^{1/3}-1)4/3 \sqrt{1-r^{2}/R_A^{2}}$. For flat disk we have
$\alpha_{A}(r)\to \bar \alpha_A = 0.1(A^{1/3}-1)$.

The upper limit for the impact parameter is $b_{\max}\simeq 0.632\, R_{Au}$ for
$10\%$ central Au+Au collisions. Neglecting initial $k_{\perp}$ in Eq.\
(\ref{eq.30}) it is convenient to introduce the new variable $x_a \equiv x_1 =
x_{\perp}^{\pi}\xi/2z$, $x_b \equiv x_2 = x_{\perp}^{\pi}\xi/2z(\xi-1)$, where
$x_{\perp}^{\pi} = 2p_{\perp}^{\pi}/\sqrt{s}$, $z = x_{\perp}^{\pi}/x_{\perp}$,
$x_{\perp} = 2E/\sqrt{s}$, i.e. $z = p_{\perp}^{\pi}/E$. After integration over
$r$ and $b$ in Eq.\ (\ref{eq.30}) we obtain cross section for hadron production
in gluon jet in the central Au+Au collision:
\begin{widetext}
\begin{eqnarray}
\label{eq.31}  E_h \frac{d\sigma_{h}^{AA}}{d^{3}p} &=&
\frac{9K}{(p_{\perp}^{\pi})^4}\int\limits_{x_{\perp}^{\pi}}^{z_{\max}}dz\,\alpha
_{s}^{2}(Q(z))
z^2   D_{h/g}(z^{*}_c,\hat {Q}^2) \frac{1}{(1-\frac{\Delta
E}{E}(z)}\times \nonumber
\\&\times&\int\limits_{\frac{2z}{2z-x_{\perp}^{\pi}}}^{\frac{2z}{x_{\perp}^{\pi}
}} d\xi f^{gg}(\xi) \Phi(x_1,x_2) x_{1}G_{g}(x_1)x_{2}G_{g}(x_2) \,,
\end{eqnarray}
\end{widetext}
where $Q = p_{\perp}^{\pi}/2z$, $\hat{Q} =
p_{\perp}^{\pi}/2z^{*}$, and $G_{g}(x_1)$, $G_{g}(x_2)$ are the
gluon structure functions. $D_{h/g}$ is fragmentation
function~\cite{18} and $K$ is K-factor. Function $\Phi(x_1,x_2)$
describes the contribution of shadowing. For spherical nuclei (for
10\% central collisions) we obtain: $\Phi(x_1,x_2) = 226.58\,
S_{0}(x_1)S_{0}(x_2) - 117.27\, S_{1}(x_1)S_{0}(x_2) - 97.69\,
S_{0}(x_1)S_{1}(x_2) + 54.87\, S_{1}(x_1)S_{1}(x_2)$. Functions
$S_{0}$ and $S_{1}$ can be found in Ref.\ \cite{17}. The value
$f_{gg}(\xi)$ is the fraction of gluon-gluon elementary
cross-section in variable $\xi$:
\begin{equation}
\label{eq.32} f_{gg}(\xi) = \frac{3(\xi-1)}{\xi^4} - \frac{(\xi-1)^2}{\xi^6} +
\frac{1}{(\xi-1)\xi^3} + \frac{(\xi-1)^2}{\xi^3}\,.
\end{equation}
We also have:
\begin{equation} \label{eq.33}
f_{qg}(\xi) =
\frac{1+\xi^2}{\xi^{4}(\xi-1)} + \frac{4(\xi-1)}{9\xi^5}
+\frac{4(\xi-1)}{9\xi^3}
\end{equation}
\begin{equation}
\label{eq.34} f_{qq}(\xi) = \frac{4(1+\xi^2)}{9\xi^{4}(\xi-1)}
\end{equation}

The value $z_{\max}$ can be found from Eq.\  (\ref{eq.29}):
\mbox{$z^{*}_c = \frac{p_{\perp}^{\pi}}{E(1 - \frac{\Delta E}{E})}
= \frac{p_{\perp}^{\pi}}{E - \Delta E} = 1$} (taking into account
that $z = p_{\perp}^{\pi}/E$). The value of $E$ which corresponds
to gluon jet with some $p_{\perp}^{\pi}$ can be found from Fig.\
\ref{Fig.1}(in large scale). For example, for values
$p_{\perp}^{\pi} =3,\, 4,\, 6$ GeV/c we can estimate $z_{\max} =
0.61,\, 0.64,\, 0.67$ (e.g.\ at $p_\perp=4$ we have $E - \Delta
E(E)\simeq 6.25-2.25 = 4$, and $z_{max}\simeq 4/6.25 = 0.64$).

We obtain the approximate numeral formulas for $\frac{\Delta E}{E}(z)$ in the
region $z_{\min}< z < z_{\max}$ for several values of $p_{\perp}^{\pi}$ (which
are $(3-5)\%$ precise). For example, for $p_{\perp}^{\pi}=4$ GeV we have (here
$z_{\min}= 2P_{\perp}/\sqrt{s} \simeq 0.0615$):
\begin{equation}
\label{eq.35} \frac{\Delta E}{E}(z) \simeq 0.46z + 0.061 +
0.95(z-0.0615)(0.64-z).
\end{equation}
The relations of this kind we use in formula (\ref{eq.31}) for gluon jets.

But it should be noted, that for the quark scattering on gluon and quark
potentials the color factor decrease 16/81 = 0.198 times and the energy loss
$\Delta E$ should decrease correspondingly. From relation $E - 0.198 \Delta E_g
= p_{\perp}^{\pi}$ we can find for $p_{\perp}^{\pi} =3,\, 4,\, 6$ GeV the
corresponding values $z_{\max}\simeq 0.91,\, 0.92, \, 0.924$. Here we also find
the approximate analytical formulas for $\frac{\Delta E}{E}(z)$, which we use
in Eq.\ (\ref{eq.31}) for quark jets. For example at $p_{\perp}^{\pi}$ = 4 GeV
we have:
\begin{equation}
\label{eq.36} \frac{\Delta E}{E}(z)\simeq 0.068z + 0.014
+0.11(z-0.0615)(0.92-z)
\end{equation}

In this work we calculate the effective suppression factor for the $\pi^0$
spectra at RHIC energy, or the ratio:
\begin{equation}
\label{eq.37} R_{AA}(p_{\perp}) = \frac{\frac{dN_{AA}}{dy
d^{2}p_{\perp}}}{\sigma_{in}^{pp} \frac{dN_{pp}}{dyd^2p_{\perp}} \bar{T}} \;,
\end{equation}
where $ \bar{T}= \int\limits_0^{b_{\max}}T(b)d^{2}b \left/
\int\limits_0^{b_{\max}}d^{2}b \right.$. This is ratio between the spectrum in
central AA and pp collisions, which is normalized to the effective total number
of binary NN collision in central AA collisions. If we do not take into account
the nuclear effects (jet quenching and shadowing), this ratio should be unity
at large $p_{\perp}$.

In Eq.\ (\ref{eq.31}) for invariant cross section we take into account  the
following parton collisions:
\begin{enumerate}
    \item The gluon-gluon collision, the gluon scattering by gluon
potential and gluon fragmentation into $\pi^0$. This process is described by
Eq.\ (\ref{eq.31}) with effective fragmentation function
$D_{\pi^{0}/g}(z^{*}_c,\hat Q^{2})$~\cite{18} and with function $\frac{\Delta
E}{E}(z)$ of the kind (\ref{eq.35}) for each value of $p_{\perp}^{\pi}$.
    \item The gluon-quark collision (with u and d quarks), the gluon
scattering by gluon and quark potentials and fragmentation of gluon. In Eq.\
(\ref{eq.31}) we use $f_{qg}(\xi)$ from (\ref{eq.33}) with corresponding
factor, the product $G_{g}G_{q}$ from \cite{12} and $\Delta E/E$ of the kind
(\ref{eq.35}).
    \item  The $qg$ collision, the quark scattering by gluon and quark
potentials and fragmentation of quark into $\pi^0$. In Eq.\ (\ref{eq.31}) we
have now $D_{\pi^{0}/q}$~\cite{18} with $\frac{\Delta E}{E}(z)$ of the kind
(\ref{eq.36}) for each value of $p_{\perp}^{\pi}$ due to the decrease of the
color factor.
    \item The quark-quark collision $f_{qq}(\xi)$ (\ref{eq.34}), the
quark scattering by gluon and quark potentials and fragmentation of quark. We
have here $\Delta E/E$ also of type (\ref{eq.36}) and product $G_{q}G_{q}$.
    \item  The collision of gluon with sea quarks, the gluon scattering by
gluon and quark potentials and gluon fragmentation. Here we also use the
effective fragmentation function $D_{\pi^{0}/g}$ with $\Delta E/E$ of the kind
(\ref{eq.35}) and corresponding structure function from Ref.\ \cite{12}.
\end{enumerate}

In order to find  the effective suppression factor $R_{AA}(p_{\perp})$ of the
$\pi^0$ spectra, the cross section (\ref{eq.31}) should be divided by the same
cross section for the same value of $p_{\perp}^{\pi^0}$, but without nuclear
effects ($\Delta E = 0, z_{max}=1$), and without shadowing
($S_{0}(x_1)=S_{0}(x_2)=1$, $S_1(x_1)=S_1(x_2)=0$ for spherical nuclei).

\begin{figure} [ht]
\centering \mbox{\includegraphics*[scale=0.8]{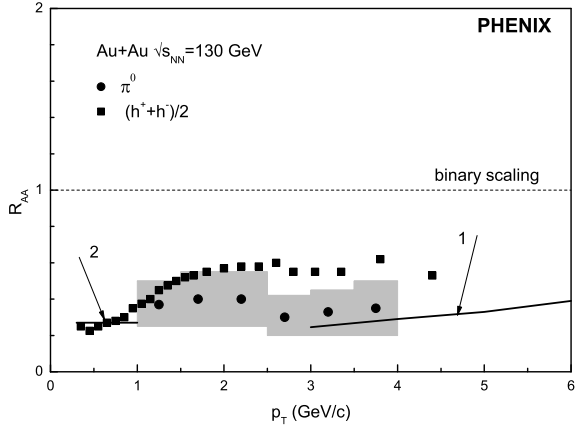}} \caption{The
suppression factor $R_{AA}(p_{\perp})$ for charged hadrons and neutral pions.
The gray band for $\pi^0$ is defined by the sums of squares of the systematic
errors of the measurement and uncertainty in the $N+N$ reference and in
$\langle N_{binary}\rangle$. The errors for $(h^{+} + h^{-})/2$ are not shown.
The line 1 is the calculated factor $R_{AA}(p_{\perp}^{\pi^0})$ for hard
processes, the line 2 is factor for soft processes calculated in assumption
that the low $p_{\perp}$ spectra scale like $A^{1.1}$. \label{Fig.3}}
\end{figure}

We have calculated such ratio for hard collisions (3 GeV/c $\le
p_{\perp}^{\pi^0} \le$ 6 GeV/c). For example, this ratio for
$p_{\perp}^{\pi^0}$ = 4 GeV/c (the contributions of collisions listed above in
items 1--5 are given in numerator and denominator) is:
\begin{eqnarray}
\label{eq.38} && R_{AA}(p_{\perp}^{\pi^0} = 4 \textrm{GeV/c}) =
\nonumber \\ &=& \frac{0.37 + 0.08 +1.09 +0.27 +0.015}{4.0 + 0.53
+ 1.8 + 0.40 +0.161} \simeq 0.27.
\end{eqnarray}
The values of $R_{AA}$ corresponding to $p_{\perp}^{\pi^0} = 3,\,
4,\, 5,\, 6$ GeV are $R_{AA} = 0.23,\, 0.27,\, 0.32,\, 0.38$. We
show in Fig.~\ref{Fig.3} the ratio $R_{Au,Au}(p_{\perp}^{\pi^0})$
and the experimental data at RHIC energy. The theoretical values
of $R_{Au,Au}$ are somewhat smaller, than the experimental values,
though they apparently are within the limits of experimental
uncertainly.

Theoretical values of $R_{AA}$ in Fig.~\ref{Fig.3} correspond to sufficiently
large values of $p_{\perp}$. In the intermediate region $1\le p_{\perp}\le 3$
GeV there are uncertainties associated with the interplay of contributions from
hard and soft processes. In the soft region $p_{\perp} < 1$ GeV/c it is
possible make the estimate of $R_{AA}$ by Eq.\ (\ref{eq.37}) if the scaling
$A^{1.1}$ is used: $\frac{dN_{AA}}{dy} \simeq A^{1.1} \frac{dN_{pp}}{dy}$. This
gives the value $R_{AA}\simeq 0.25-0.29$, which is close to experimental data
available for charged particles.

\section{Conclusion \label{sec4}}

In this paper we investigate the energy loss of quark and gluon
jets in the quark-gluon plasma and suppression of the $\pi^0$
spectra at RHIC energy. We use the initial conditions in plasma,
which were found earlier in the effective quasiparticle model for
SPS and RHIC energies \cite{5}. We take into account also the
possibility of production of the hot glue at the first stage with
more high temperature and density. It can be shown, that plasma is
sufficiently thin: $\bar n =
\int\limits_{\tau_{g}}^{\tau_{c}}d\tau\;
n_{g}(\tau)\sigma_{gg}(\tau) \simeq 1.28$, where $\bar n$ is an
average number of jet scatterings. The energy loss in  expanding
plasma is calculated in the dominant first order \cite{3,4} taking
into account the finite kinematic limits. The energy losses
$\Delta E$ of gluon and quark jets are calculated in a wide range
of parton energies $E$. These energy losses are used for
calculation of suppression of $\pi^0$ spectra with moderately high
$p_{\perp}$ ($3 < p_{\perp} < 6$ GeV/c), which is done by
modification of gluon and quark fragmentation functions. We also
take into account the parton shadowing factors in nucleus
\cite{17}. We obtain a considerable suppression of $\pi^0$
spectra, which is caused by the effects mentioned above. This
suppression agrees with the data reported by PHENIX in the region
$p_{\perp}^{\pi_0} = 3 - 4$ GeV/c. The estimates of $R_{AA}$
(\ref{eq.37}) in the soft region ($p_{\perp} < 1$ GeV/c) also
agrees with experimental data if the scaling $A^{1.1}$ is taken
into account.

The intrinsic transverse momentum $k_{\perp}$ at RHIC energy is neglected in
this work. It is important at SPS energy (Cronin effect), but at RHIC energy
one can neglect the effects of initial $k_{\perp}$ with a good precision
\cite{16}. In principle a small increase of $R_{AA}$ is possible at moderate
$p_{\perp}$ because of transverse momentum broadening. We are planning to take
into account the intrinsic $k_{\perp}$ in the next work.

The most important conclusion of this work is the possibility to investigate
the structure of the plasma phase transition into hadrons with the help of hard
processes. The correct quantitative description of $\pi^0$ suppression is
probably possible only in a model of phase transition which includes the
decrease of thermal gluon mass $m_{g}(T)$ and effective coupling $G(T)$ near
the phase transition ($T \simeq T_c$) point. However, the quasiparticle model
with increase of these values at $T \to T_c$ leads to a too high energy losses
(which can even exceed the energy of the jet itself). The main reason of this
problem is too large value of the coupling $G(T)$ near $T_c$. We show also that
energy losses at SPS energy are very small as a consequence of a too low
initial temperature of the plasma phase at this energy.

It is interesting to investigate also the suppression of charged meson and
baryon spectra, which we are planning to do in the forthcoming paper.

\begin{acknowledgements}
The author thank Profs.\ S.\ T.\ Beljaev, V.\ I.\ Manko and
especially S.\ L.\ Fokin for fruitful discussions. The author is
grateful to Prof.\ B.\ V.\ Danilin, Drs.\ A.\ Lomonosov, and \ L.\
V.\ Grigorenko for careful reading of the manuscript. The work was
supported by the grant NS-1885.2003.2 of the Russian Ministry of
Industry and Science.
\end{acknowledgements}

\end{document}